\pdfoutput=1
\documentclass[11pt]{article}

\usepackage[]{emnlp2021}

\usepackage{times}
\usepackage{latexsym}
\usepackage{inconsolata}
\usepackage{refstyle}
\usepackage{multirow}
\usepackage{times}
\usepackage{latexsym}
\usepackage[utf8]{inputenc}
\usepackage{xcolor,colortbl}
\usepackage{microtype}
\usepackage{url}
\usepackage{longtable}
\usepackage{tikz}
\usepackage{tabularx}
\usepackage{lscape}
\usepackage{hhline}
\usepackage{bbm}
\usepackage{array}
\usepackage{xspace}
\usepackage{graphicx}
\usepackage{booktabs}
\usepackage{makecell}
\usepackage{amsfonts}       % blackboard math symbols
\usepackage{nicefrac}       % compact symbols for 1/2, etc.
\usepackage{microtype}      % microtypography
\usepackage{multirow}
\usepackage{caption}
\usepackage{amsmath}
\usepackage{float} 
\usepackage{ragged2e}
\usepackage{breqn}
\usepackage{soul}
\usepackage[T1]{fontenc}
\newcommand{\modelname}{\textsc{RETAIN}\xspace}
\usepackage[utf8]{inputenc}

\usepackage{microtype}

\usepackage[colorinlistoftodos,textsize=tiny]{todonotes}
\usepackage{hyperref}

\title{RETAIN: Interactive Tool for \\ Regression Testing Guided LLM Migration}
\author{Tanay Dixit$^{1}$ \thanks{\;\;This work was done when TD was interning at Adobe.} ~~~ Daniel Lee$^{2}$  ~~~ Sally Fang$^{2}$  ~~~ \textbf{Sai Sree Harsha$^{2}$}  \\ ~~~ \textbf{Anirudh Sureshan$^{2}$}  ~~~ \textbf{Akash Maharaj$^{2}$}   ~~~ \textbf{Yunyao Li$^{2}$}  \\
$^{1}$University of Illinois Urbana Champaign \\
$^{2}$Adobe Inc.\\
\texttt{\{tdixit, dlee1\}@adobe.com}
}

\begin{document}
\maketitle

\begin{abstract}
%Large Language Models (LLMs) are being integrated into many applications \cite{kaddour2023challenges}. However, the rapidly evolving LLM space requires developers to constantly adapt, leading to potential performance regressions when migrating LLMs. To reduce the workload of prompt engineering several interactive tools have been proposed, but very few cater to the specific needs of regression testing for prompting LLMs \cite{ma2024my}. To fill this gap we propose \modelname - designed specifically for regression testing for prompting LLMs. We build an interactive tool catered to the needs of regression testing, and develop an error discovery module in order to help users in understanding differences in model behaviours. The error discovery module provides textual descriptions of the various errors or difference between model outputs, which leads to actionable for making prompt updates. Through empirical evaluations and a comparative study we show that \modelname, when compared to manual evaluation, helped participants identify twice as many errors, experiment with Y\% more prompt versions, and reach satisfactory prompts with X\% time. \footnote{Tool available \url{https://github.com/adobe/starter-repo} under Apache V2 license.}%

Large Language Models (LLMs) are increasingly integrated into diverse applications \cite{kaddour2023challenges}. The rapid evolution of LLMs presents opportunities for developers to enhance applications continuously. However, this constant adaptation can also lead to performance regressions during model migrations. While several interactive tools have been proposed to streamline the complexity of prompt engineering, few address the specific requirements of regression testing for LLM Migrations \cite{ma2024my}. To bridge this gap, we introduce \modelname (REgression Testing guided LLM migrAtIoN), a tool designed explicitly for regression testing in LLM Migrations. \modelname comprises two key components: an interactive interface tailored to regression testing needs during LLM migrations, and an error discovery module that facilitates understanding of differences in model behaviors. The error discovery module generates textual descriptions of various errors or differences between model outputs, providing actionable insights for prompt refinement. Our automatic evaluation and empirical user studies demonstrate that \modelname, when compared to manual evaluation, enabled participants to identify twice as many errors, facilitated experimentation with 75\% more prompts, and achieves 12\% higher metric scores in a given time frame. %These results underscore \modelname's efficacy in streamlining the LLM regression testing process, potentially accelerating development cycles and improving prompt quality.

\end{abstract}
\section{Introduction}
%Large Language Models have shown to excel at performing several complex tasks \cite{achiam2023gpt, Dubey2024TheL3} which would previously required custom fine-tuned models. As a results using LLMs in applications become highly lucrative \cite{}, as it saves costs in developing models from scratch. For LLMs to effectively perform complex tasks, it requires carefully designing prompts \cite{wei2022chain, brown2020language}. Prompt engineering is an unstructured task involving carefully defining the instructions in the prompt or curating a set of in context examples \cite{khattab2022demonstrate}. These design choices are often are highly specific to the model being prompted. %

Large Language Models (LLMs) have demonstrated proficiency in executing a wide array of complex tasks \cite{achiam2023gpt, Dubey2024TheL3}, which previously necessitated custom fine-tuned models. This capability has made the integration of LLMs into applications increasingly attractive, as it significantly reduces the costs associated with developing models from scratch. However, for LLMs to effectively perform these complex tasks, careful prompt design is crucial \cite{brown2020language, wei2022chain}. Prompt engineering is an unstructured process that involves crafting instructions within the prompt or curating a set of in-context examples \cite{khattab2022demonstrate}. These design choices are often highly specific to the particular model being prompted.
%, requiring a nuanced understanding of each model's idiosyncrasies and capabilities.
\\

\begin{figure}[t]
    \centering
    \includegraphics[width=0.8\linewidth]{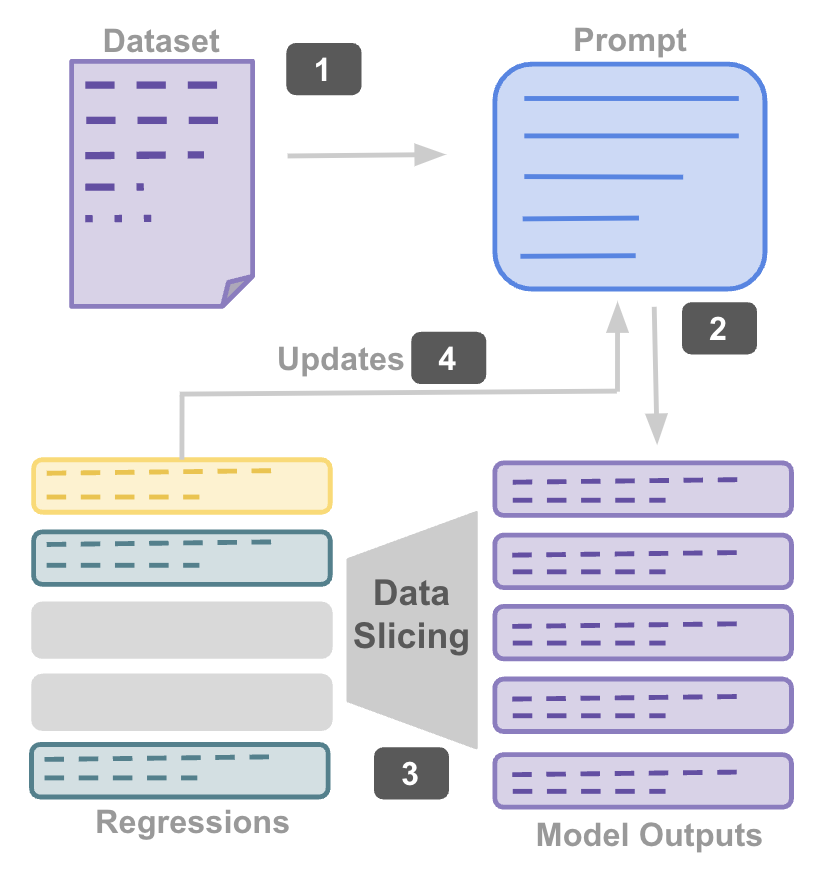}
    \caption{Regression Testing for Prompting LLMs. The process involves: (1) input dataset, (2) initial prompt, (3) data slicing algorithm to identify behavioral differences (regressions) across models, and (4) prompt refinement to address identified regressions.}
    \label{fig:enter-label}
\end{figure}

%The rapidly evolving landscape of LLMs forces application developers to continually update to newer versions to maintain optimal performance. Additionally, applications using LLM APIs often get forced to switch the models as LLMs get deprecated and discontinued\footnote{https://platform.openai.com/docs/deprecations}. This creates a repeated effort of prompt engineering a different LLMs for the same task. \citet{ma2024my} show that migrations to newer LLMs are challenging due to model regressions, requiring researchers to build custom tools for analyzing differences in model behaviors. %

The rapidly evolving landscape of LLMs, compels application developers to continually update to newer versions to maintain optimal performance. Moreover, applications utilizing LLM APIs often face forced transitions as older models are deprecated and discontinued\footnote{https://platform.openai.com/docs/deprecations}. This creates a recurring challenge of re-engineering prompts for different LLMs to achieve the same task and maintain consistent model behavior, a process we define as \textbf{LLM migration}.

Migrations to newer LLMs are difficult due to model regressions \cite{ma2024my}, necessitating the development of custom tools for analyzing discrepancies in model behaviors. Such regression tests must focus both on pattern discovery for errors and systematic failure validation \cite{cabrera2023zeno, ma2024my}. These patterns can generally be encoded as a subgroup or ``slice" of model outputs, with a corresponding metric that characterizes the observed behavior, and are often discovered in an iterative and manual manner by prompt developers \cite{shankar2024spade}.

Figure~\ref{fig:enter-label} illustrates a high-level regression testing process for prompting, drawing parallels with software engineering techniques. The main challenge in regression testing based prompting, is to design a systematic method of identifying regressions. While numerous tools and frameworks have been developed to assist in prompt engineering, ranging from interactive platforms \cite{wu2022promptchainer, arawjo2024chainforge, cabrera2023zeno} to automated systems \cite{khattab2023dspy, zhou2022large}, few address the specific needs of regression testing in prompting. Existing tools often lack support for data slicing (Figure~\ref{fig:enter-label}), which requires manual inspection to identify regressions and group data points into slices. Furthermore, current tools provide insufficient support for analyzing model behaviors at various granularities. 
% \vspace{-0.5pt}
\par To bridge this gap, we propose \modelname (\textbf{RE}gression \textbf{T}esting guided LLM migr\textbf{A}t\textbf{I}o\textbf{N}) - designed explicitly for regression testing in prompting and enables flexible analysis of model behaviors at various granularities. \modelname aims to reduce the effort required in identifying regressions by automatically detecting differences in model behaviors across different data subsets (\S\ref{subsec:error_panel}). Our tool features an interactive interface supporting the analysis of various prompt iterations across multiple granularity levels: aggregate metric scores, distribution analysis of metric scores, and side-by-side comparisons at the instance level (\S\ref{subsec:data_panel}). Furthermore, \modelname integrates prompt updating capabilities, making it a self-contained solution for the entire prompting process. Through user studies, we demonstrate that \modelname, compared to manual prompting approaches, aids users in identifying twice as many errors, facilitates iteration over 75\% more prompts, and achieves 12\% higher metric scores in a given time frame. 

\section{Related Work}
\subsection{Prompting tools}
Prompting has emerged as new paradigm \cite{liu2023pre} based on language models that model the probability of text directly. To effectively leverage the pre-trained knowledge of large language models (LLMs), carefully designed prompts are required \cite{wei2022chain}. To facilitate analzing and experimenting with different prompt several commercial prompting tools and libraries, such as Promptify \cite{Promptify2022}, Lang Chain \cite{langchain} and Guidance \cite{guidance} have been developed. Several interactive prompting tools like \citet{strobelt2022interactive, mishra2023promptaid, wu2022promptchainer} aim to reduce the workload in experimenting with several prompts. Tools like Zeno \cite{cabrera2023zeno} provide support for analysing models performance on different data slices but are limited to only datasets that contain meta-data, which is often not available for majority NLP tasks. A new emergent area involves automatic prompt engineering techniques \cite{khattab2023dspy, yuksekgonul2024textgrad} which aim to treat the prompting process as an optimization task. 

\subsection{Exploratory Analysis and Automated Discovery}
\begin{figure*}[t!]
    \centering
    \includegraphics[width=0.9\linewidth]{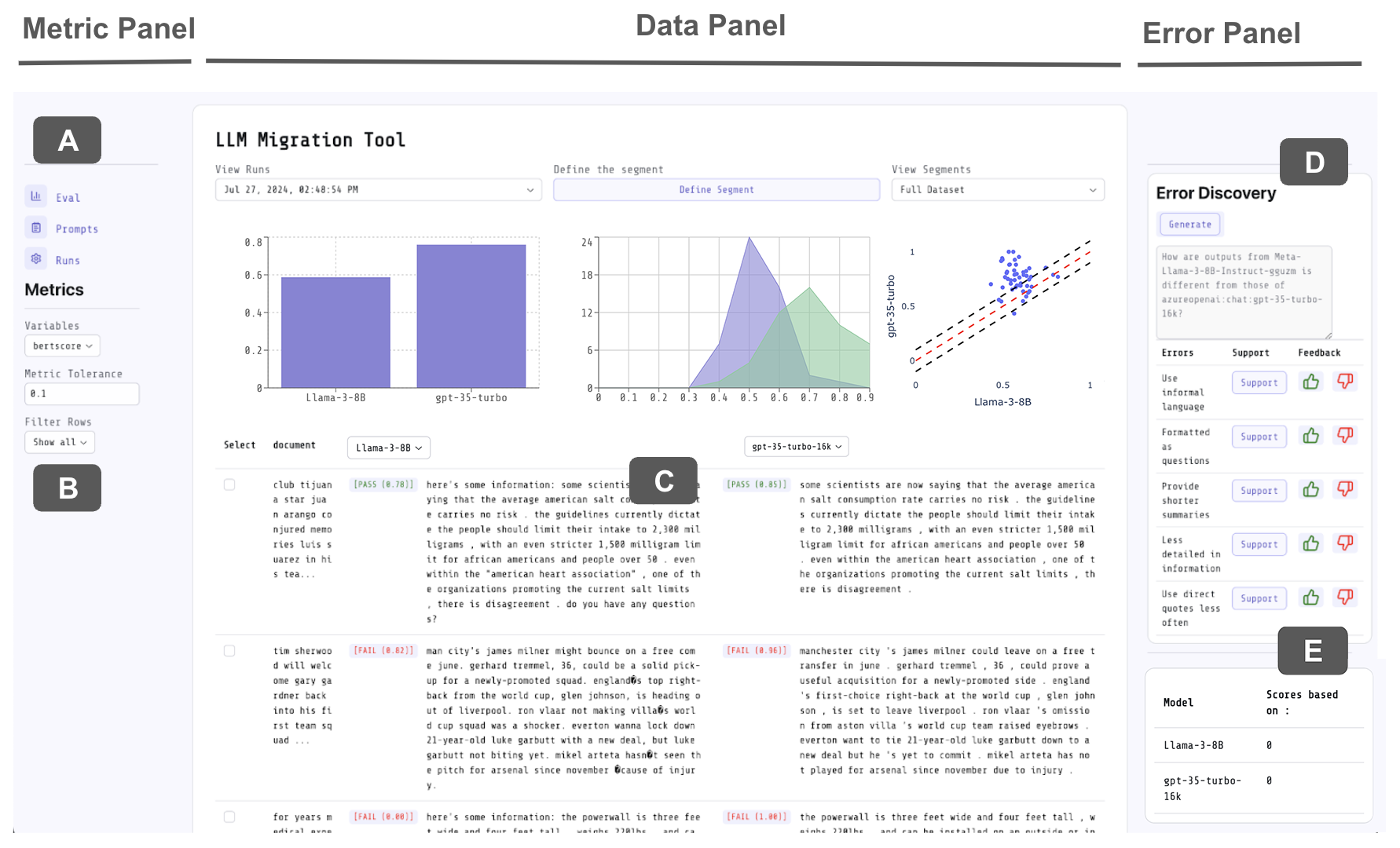}
    \caption{\modelname comprises of three main Panels: Metric Panel, Data Panel, and Error Analysis Panel. It features three pages (A) designed for various prompt engineering tasks (B) Users can set metrics, (C) compare model outputs through charts and side-by-side comparisons, and (D) conduct in-depth analysis of failure cases using the error discovery module. Additionally, users can define LLM assertions to evaluate outputs across different prompts. (E)}
    \label{fig:main_fig}
\end{figure*}

Automatic pattern discovery is a well studied problem with several classical methods in ML \cite{manning1999foundations} such as topic modeling \cite{10.5555/944919.944937} to extract major topical variations. Our task is different from these traditional settings as it requires error discoveries in the form of natural language predicates, which are interpretable and can express abstract concepts. Several works like  \citet{NEURIPS2023_7e810b2c, wang2023goal, zhong2022describing} show that LLMs are capable of extracting distributional differences between two text corpora. We leverage these ideas for building our data slicing module  (Figure~\ref{fig:main_fig}-D).

\section{User Challenges in Regression Testing for LLM Migrations}
%To understand practitioners’ workflows in performing regressions testing for prompting LLMs, we had conversations with researchers and engineers. We learned that they found it challenging to identify differences in model outputs (regressions) and understand the causes of variations in metric scores. Additionally, they struggled to systematically track the results of prompt edits on model outputs. These issues particularly arised in scenarios involving prompt migrations. Prompt migrations refer to transitioning prompts for a given task from one LLM to another. In such cases, it is critical to ensure that the behavior of the LLM remains consistent, making regression testing essential. We summarize the key design goals as follows - %

To understand users' workflows in regression testing for LLM Migrations, we conducted a formative study and collaborative design process, adapted from the methodology described in \cite{10.1145/3491102.3517479}. Our study included semi-structured interviews with researchers and engineers, focusing on their experiences in LLM Migrations. 
%The collaborative design component worked to identify areas of user pain points within the participants' past experiences and explored their current projects requiring regression testing and prompt engineering.

Our findings revealed several key challenges: difficulty in identifying differences in model outputs (regressions), struggle to understand causes of variations in metric scores, and lack of systematic tracking for the effects of prompt edits on model outputs. In cases of migrations, ensuring consistent LLM behavior is critical, underscoring the importance of regression testing. Based on these insights, we identified three primary design goals:

\begin{itemize}
\item \textbf{DG1:} Develop methods to automatically identify behavioral changes across prompts or models, and intelligently suggest data slices, especially when metadata is unavailable. 
\item \textbf{DG2:} Provide tools for examining LLM behavior at various levels, from aggregate metrics to individual instance comparisons, supporting diverse analytical needs.
\item \textbf{DG3:} Integrate capabilities for systematic tracking and analysis of prompt modifications, enabling users to iterate and improve prompts based on regression testing results.
\end{itemize}

\section{System}

%In this section we walk through an example of an ML practitioner, who is using our tool for regression testing based prompting. We focus on the task of prompt migration \cite{ma2024my} for a task of abstractive summarization \cite{DBLP:conf/nips/HermannKGEKSB15}, but the tool can be used for any prompting engineering setup. The user first starts of with a simple declarative config file (Appendix~\S\ref{app:imp_details}) which contains the models names, access keys, initial prompts, metrics and test data \cite{promptfoo}. Using this the user can spin up the \modelname tool. Tool implementation details in \ref{app:imp_details}.%

In this section, we demonstrate \modelname using a scenario where a researcher or engineer utilizes our tool for LLM migration in the task of prompt migration \cite{ma2024my} for 
 a summarization task \cite{DBLP:conf/nips/HermannKGEKSB15}. The user is migrating a prompt optimized for \texttt{gpt3.5-turbo-16k} to \texttt{llama-3-8b}. It's important to note that \modelname is versatile and applicable to any prompt engineering setup.
The user initiates the process by creating a simple declarative configuration file (detailed in Appendix~\S\ref{app:imp_details}). This file contains essential information such as model names, access keys, initial prompts, metrics, and test data \cite{promptfoo}. With this configuration in place, the user can launch the \modelname tool. For implementation details, readers are directed to Appendix~\ref{app:imp_details}.

%\subsection{Pages} The Eval Page which contains three panels - (i) Metric Panel (ii) Data Panel (iii) Error Analysis Panel . The Prompts page contains the model's prompt - here it's the \texttt{Llama-3-8B} model prompt (Appendix~\S\ref{app:imp_details}). For the given task of migration the user starts of with the same prompt as \texttt{gpt3.5-turbo} and make edits to optimize the Llama prompt to have the same behaviour as \texttt{gpt3.5-turbo}. Finally, the Runs page presents a tabular view of the metrics scores of the two models (Appendix~\S\ref{app:imp_details}). Aims to provide a bird eyes view of the prompt engineering process. 

\subsection{Pages}\modelname consists of three tabs: (1) Eval, (2) Prompts, and (3) Runs (Figure~\ref{fig:main_fig}-A). The Eval Page comprises three key panels: (i) Metric Panel, (ii) Data Panel, and (iii) Error Analysis Panel. The Prompts page (Figure~\ref{fig:prompts_page}) displays the model's prompt, which in this case is the prompt for \texttt{Llama-3-8B} model.
For the task of migration, the user begins with the same prompt as \texttt{gpt3.5-turbo} and iteratively refines it to optimize the Llama prompt, aiming to achieve behavior comparable to \texttt{gpt3.5-turbo}. The Runs page (Figure~\ref{fig:runs_page}) offers a tabular view of the metric scores for both models. This structure is designed to provide a comprehensive overview of the prompt engineering process, offering users a bird's-eye view of the entire migration workflow. 
%It enables efficient comparison and analysis of model performance, facilitating informed decision-making in the prompt optimization process.

%\subsection{Metrics Panel} All the metrics the defined by the user in the config file would be displayed in the variables toggle in the Metrics Card (Figure~\ref{fig:main_fig}-B). To account for non-determinism in LLM regression testing \cite{ma2024my} we include - Metric Tolerance - analogous to confidence interval in hypothesis testing. This represents the acceptable difference between two metric scores for them to be considered equivalent. The last drop-down option is two filter the data table to only display the test data points where the metric scores differs by more than the set tolerance.

\subsection{Metrics Panel} The Metrics Panel displays all user-defined metrics from the configuration file within the Metrics Card's variables toggle (Figure~\ref{fig:main_fig}-B). To address the challenge of non-determinism in LLM regression testing \cite{ma2024my}, we introduce the concept of Metric Tolerance. This feature is analogous to confidence intervals in hypothesis testing and represents the acceptable margin of difference between two metric scores for them to be considered equivalent. The panel features a dropdown menu for filtering the data table to display only test data points where metric score differences exceed the set tolerance. This enables users to focus on discrepancies between model outputs, aiding in efficient analysis and debugging.

%\subsection{Data Panel}
%The Data Panel (Figure~\ref{fig:main_fig}-C) consists of aggregate-level visualizations and instance-level side-by-side comparisons.

%\paragraph{Visualizations} The user can compare the performance of the model across three different visualizations (1) Aggregate Metric Score chart. Since aggregate scores do not reflect the true model behaviour \cite{cabrera2023zeno, ribeiro2020beyond}, the user can also make use of (2) Metric score distribution chart - analyzes whether the model has the same metric score distribution as the target. Catering specifically to our design goal of regression based prompting, the user can also visualize regressions via (3) Regressions chart - plots the given metric score with the migration model on Y axis and base model on X axis, all data points that regress would lie outside the black lines. \\

%\paragraph{Side-by-Side Comparisons} The user can also inspect the model outputs at the instance level through the side-by-side tabular comparisons. Providing such instance level comparisons is critical for users to identify their slices of interests and to observe qualitative patterns \cite{kahng2024llm}.

\subsection{Data Panel}
\label{subsec:data_panel}
The Data Panel (Figure~\ref{fig:main_fig}-C) consists of aggregate-level visualizations and instance-level side-by-side comparisons (DG2)

\paragraph{Visualizations}
The panel incorporates three visualizations to facilitate model analysis. First, the Aggregate Metric Score Chart provides a performance summary. However, recognizing that aggregate scores may not fully capture model behavior \citep{cabrera2023zeno, ribeiro2020beyond}, we include additional visualizations. The Metric Score Distribution Chart allows users to compare the distribution of metric scores between the models. Lastly, the Regressions Chart \cite{promptfoo}, designed to address our goal of regression-based prompting.

%plots the given metric scores with the migration model on the Y-axis and the base model on the X-axis. Data points falling outside the diagonal indicate instances of regression, enabling quick identification of performance changes.

\paragraph{Side-by-Side Comparisons}
To complement the aggregate visualizations, we provide instance-level comparisons through a side-by-side tabular interface. This feature is crucial to identify specific slices of interest and observe qualitative patterns in model outputs \citep{kahng2024llm}. By allowing direct comparison of individual instances, users can gain deeper insights into the model's behavior. % across various input types and characteristics.

%\subsection{Error Analysis Panel} A major time painpoint in prompt engineering is identifying where and why the model is scoring poorly on given metrics. To help users identify errors quickly and to facilitate action drive prompt edits, we introduce - Goal driven error discovery. 

\begin{figure}[h!]
    \centering
    \includegraphics[width=0.9\linewidth]{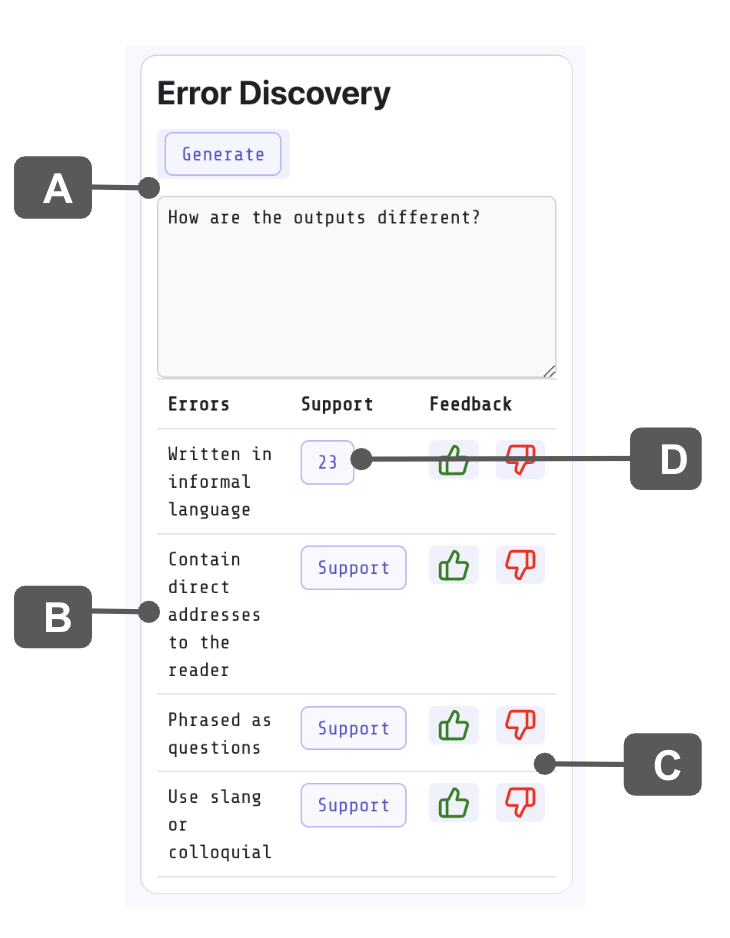}
    \caption{Error Discovery Module Interaction. (A) Users initiate error generation to identify discrepancies among model outputs in the side-by-side comparison table. (B) For errors of interest, users can employ the support feature (D) to highlight specific model outputs containing the selected error type. (C) The thumbs up/down feature allows users to create or remove custom LLM metrics based on error descriptions.}
    \label{fig:error_disc}
\end{figure}

%\paragraph{Goal Driven Error Discovery}
%The error discovery module is inspired from \cite{zhong2022describing, NEURIPS2023_7e810b2c}. It aim to identify distributional difference between the model outputs that are relevant to given user-defined goal. This helps not only user understand why the the model is scoring poorly on the given metrics but also provides textual descriptions of the errors which can be incorporated in subsequent prompt edits. Upon user request we also identify which specific model outputs contain which errors through a validator module. 

%We define two pipelines for performing these tasks. For the goal oriented error discovery we prompt GPT-4 to identify differences between the two model outputs \cite{NEURIPS2023_7e810b2c}. Prompt present in Appendix~\ref{app:error_discovery}. We limit to only displaying 5 error types. 

%\paragraph{Defining LLM Assertions}
%\citet{shankar2024spade} highlight the importance of LLM assertions to catch data quality errors made by LLMs. Following \citet{shankar2024spade, zheng2024judging}, we allow users to define their own LLM based metrics that can specifically evaluate a particular error of interest. By clicking on the thumbs up icon (Figure~\ref{fig:error_disc}-C) the user defines a LLM metric so specifically evaluate the errors corresponding to the textual description mentioned. The thumbs down deletes the given LLM metric. We make use of the error description while defining the LLM metric, additional details \& prompt in Appendix~\ref{app:imp_details}.

\subsection{Error Analysis Panel}
\label{subsec:error_panel}
A significant challenge in prompt engineering is understand why and where the model performs poorly with respect to the given metrics (DG1).
%efficiently identifying the sources and nature of model performance issues with respect to given metrics. 
To address this, we introduce Goal-driven error discovery, designed to streamline the error identification process and facilitate targeted prompt refinements.

\paragraph{Goal-Driven Error Discovery}
Figure~\ref{fig:error_disc} shows the various interactions with the module. Our error discovery module, inspired from \citet{zhong2022describing} and \citet{NEURIPS2023_7e810b2c}, aims to identify distributional differences between model outputs that are relevant to user-defined goals. This approach not only helps users understand why the model is under performing on given metrics but also provides textual descriptions of errors, which can be directly incorporated into subsequent prompt edits. To help users identify the model outputs containing a given error type, we employ a \textit{selector module}. The \textit{selector module} highlights the model outputs containing the specific error in the side-by-side comparison tables. We implement two distinct pipelines for these tasks. For building the goal-oriented error discovery, we prompt (Table~\ref{tab:generator_prompt}) GPT-4 to identify differences between the groups of outputs of the two models for a given goal. For the \textit{selector module} for every model output, we prompt (Table~\ref{tab:eval_prompt}) GPT-3.5 to classify whether the outputs contains the given error or not. Additional implementation details in Appendix~\ref{app:error_discovery}.

\paragraph{Defining LLM Assertions}
\citet{shankar2024spade} emphasize the importance of LLM assertions in detecting data quality errors made by language models. Building on this concept and \citet{zheng2024judging}, we enable users to define custom LLM-based metrics that specifically evaluate errors of interest. Users can create these metrics by clicking on the thumbs-up icon (Figure~\ref{fig:error_disc}-C) associated with a particular error description. In formulating these metrics, we incorporate the error descriptions to ensure relevance and specificity. Additional implementation details and we adopt the prompts from \citet{10.1145/3613904.3642216} for this task.

%\subsection{Additional features} 
%Owing to the iterative nature of prompt engineering, we support prompt versioning  - View Runs in the Data Panel (Figure~\ref{fig:main_fig}). In the formative studies we observed that practitioners were keen on computing metrics by slices to identify which slices underperform or outperform relative to others, hence to support this we provide the option to define slices (Figure~\ref{fig:main_fig}- Data Panel).Additionally, users have the flexibility to control which columns in the side-by-side comparison tables display which specific model outputs. This capability enables users to compare model outputs across different prompts. This feature is particularly advantageous when combined with the error discovery module, as it not only facilitates the comparison and contrast of how outputs change with different prompts but also leverages the error discovery module to identify differences in outputs across prompts, thereby allowing for a detailed analysis of how prompt edits impact model behavior. \\

\subsection{Features for Iterative Prompt Engineering}
To support the iterative nature of prompt engineering (DG3), we offer several additional features. The \textit{View Runs} feature in the Data Panel (Figure~\ref{fig:main_fig}) enables users to track and compare performance across different prompt versions. The \textit{Define Sgements} feature helps users define custom data slices and persist them across runs (DG2), addressing the need for fine-grained performance analysis identified in our formative studies. Users can customize which model outputs are displayed in the side-by-side comparison tables. This feature, combined with the error discovery module, allows for detailed analysis of how prompt edits affect model behavior across subgroups of data for different versions. 

\section{Evaluations}
%We conduct a small-scale empirical evaluation of our LLM-based error discovery approach to understand how accurate it is at detecting distributional differences in model outputs (\S\ref{sec:auto_eval}), and to understand how \modelname affects the prompt migration process when compared to the designers’ current practice, we conducted a within-subjects study - \S\ref{sec:user_study}.

To evaluate our system comprehensively, we employ two approaches: (1) an automatic evaluation (\S\ref{sec:auto_eval}) to assess the accuracy of our LLM-based error discovery method in detecting distributional differences between model outputs, and (2) a user study (\S\ref{sec:user_study}) to compare \modelname's impact on the prompt migration process against current practices. 
%These evaluations collectively provide insights into both the technical effectiveness of our error identification mechanism and the practical benefits of our system in real-world prompt engineering scenarios.

\subsection{Automatic Evaluations}
\label{sec:auto_eval}
%The core contribution of \modelname is leverage the goal oriented error discovery module to ease the task of identifying how and where model outputs differ. Automatically diagnosing a goal oriented error discovery system its extremely challenging, as their exists no labelled data and error discovery is an unsupervised learning task. Hence, we build a synthetic dataset to evaluate whether the system can recover known differences between two synthetic corpora.

The goal-oriented error discovery module is designed to streamline the identification of differences between model outputs. Evaluating such a system poses significant challenges due to the unsupervised nature of error discovery and the absence of labeled data. To address this, we develop a synthetic dataset to assess the system's ability to recover known differences between two artificially constructed corpora. This approach allows us to quantitatively evaluate the effectiveness of our error discovery mechanism in a controlled setting.

%\paragraph{Dataset \& Metrics}
%To synthesize a dataset to evaluate the error discovery module, we used a LLM to generate two corpora (Corpus A/B) that simultaneously differ on two different dimensions, one of which is goal-relevant and one of which is a distractor dimensions similar to the process followed by \citet{NEURIPS2023_7e810b2c}. To synthesize an example, we randomly sample a goal relevant and distractor dimension from a defined set of attributes {topic, writing style, stance, language, formatting, country}. We then synthesized Corpus A/B such that all its samples are generated using the distractor dimension while V percent of generations also use the goal-relevant dimension, where we varied V from 0.6 to 1. We synthesized 100 problems in total to create our evaluation dataset. We adopt the same metrics as \citet{NEURIPS2023_7e810b2c} to evaluate the goal oriented error discovery module, i.e., relevance to evaluate the effectivness of the module at generating errors relevant to the gold error type. To evaluate the effectivness of the module at identifying which data points in the corpora have the given error type, we use a Precision/Recall based metrics. Additional details in Appendix~\ref{app:imp_details}.

\subsubsection{Dataset Generation and Metrics}
To evaluate the error discovery module, we follow a methodology similar to \citet{NEURIPS2023_7e810b2c}. We employed a LLM to generate two corpora (A and B) that differ along two dimensions: a goal-relevant dimension and a distractor dimension. The process involves randomly sampling both dimensions from a predefined set of attributes. Corpus A and B were generated such that all samples incorporated the distractor dimension, while a fixed percent of the samples also incorporated the goal-relevant dimension.  We synthesized 100 test data points to create our evaluation dataset. For evaluation, we adopted the metrics used by \citet{NEURIPS2023_7e810b2c}. We used Error Relevance to assess the module's effectiveness in generating errors relevant to the gold error type. To evaluate the selector module (Error Coverage), we employed precision and recall metrics to evaluate the module's ability to identify data points in the corpora containing the given error type.
%This approach allows us to quantitatively assess the performance of our goal-oriented error discovery module in a controlled setting. Additional implementation details are provided in Appendix~\ref{app:imp_details}.

%\paragraph{Results}
%We analyse the performance of our error discovery module on the synthetic dataset created. We analyse the performance along two dimensions (i) How accurately the module can detect relevant errors (ii) How accurately the module can identify data points that have a given error. Table~\ref{tab:results_1} contains the results. We can see that using goals significantly improve the performance on detecting relevant errors. We also observe that the system exhibits higher precision (0.69) than recall (0.38) in identifying data points with a given error. Higher precision is advantageous as it ensures that only the rows that certainly contain the error are highlighted, thereby minimizing the need for users to manually inspect numerous rows. \\

\subsubsection{Performance Analysis} 
%We evaluated our error discovery module's performance on the synthetic dataset along two key dimensions: (i) accuracy in detecting relevant errors, and (ii) precision in identifying data points containing a given error. The results are presented in Table~\ref{tab:results_1}.
Table~\ref{tab:results_1} shows how the goal-oriented error discovery module significantly enhances the detection of relevant errors, compared with a baseline prompting  approach (see Appendix \ref{app:error_discovery} for details).  
Regarding the identification of data points with specific errors, the system demonstrates higher precision (0.69) compared to recall (0.38). This higher precision is particularly beneficial in our context, as it ensures that the system highlights rows that are highly likely to contain the error in question, reducing the burden on users by minimizing the number of rows requiring manual inspection.
% These findings suggest that our goal-oriented approach effectively improves error discovery accuracy while maintaining a user-friendly balance between precision and recall in error identification.

\begin{table}[h!]
    \centering
    \begin{tabular}{lcc}
    \toprule
    \bf Metric & \bf w/ goal & \bf w/o goal \\
    \midrule
     Error Relevance  & 0.87 & 0.72  \\
     % \citet{NEURIPS2023_7e810b2c} Error Relevance & & \\
     Error Coverage  & & \\
     ~~~~ - Precision & 0.69 & 0.70 \\
     ~~~~ - Recall & 0.38 & 0.36 \\
    \bottomrule
    \end{tabular}
    \caption{Performance Evaluation of Goal-Oriented Error Discovery. The incorporation of user-defined goals substantially enhances the accuracy of error detection, demonstrating the efficacy of our approach in identifying relevant discrepancies between model outputs.}
    \label{tab:results_1}
\end{table}

\subsection{User Study}
\label{sec:user_study}

%To evaluate \modelname, we conducted a comprehensive user study divided into two phases. This study was designed to assess two critical aspects of \modelname: sensemaking, to understand variations between different models, and prompt migration, to evaluate the end-to-end experience of migrating an LLM's performance to another.

To evaluate \modelname, we conducted a comprehensive two-phase user study designed to assess two critical aspects of our system across 12 participants proficient in prompt engineering. This dual-phase approach allows us to examine both the analytical capabilities of \modelname and its practical application in real-world prompt engineering scenarios.

\subsubsection{Phase 1: Error Identification Task}
\label{error_discovery_setup}

Phase 1 involved a within-subject study on error identification. Participants had 15 minutes per set to identify and note types of errors between two model outputs. We created a dataset with manually injected errors based on a typical LLM error taxonomy, validated by two independent NLP experts. Participants used both manual (Excel) and \modelname-assisted methods for error identification. This design compared \modelname's efficiency and accuracy against traditional methods in detecting and categorizing LLM output discrepancies, aiming to evaluate our system's potential improvements in error detection and classification.

\subsubsection{Phase 2: End-to-End Prompt Engineering Experience}

Phase 2 used a between-subject design to evaluate prompt engineering, focusing on performing LLM Migrations. Participants had 15 minutes to migrate a prompt optimized for \texttt{gpt-35-turbo} to \texttt{llama-3-8b}. Group A used a standard jupyter notebook, while Group B used \modelname, allowing comparison of \modelname's effectiveness against traditional methods in prompt engineering. After exploring \modelname, participants completed a post-screen survey using a 5-point Likert scale to assess usability, functionality, utility, cognitive load, and overall satisfaction.

\subsubsection{Results}

\modelname significantly outperformed traditional methods in regression testing guided prompt engineering. It identified nearly twice as many errors (165 vs 86) and covered more error categories (2.56 vs 2.22 average). \modelname-refined prompts achieved higher BERT scores (0.704 vs 0.625) (Figure~\ref{fig:user_results}), improving scores by 25\% compared to 12\% manually within the given timeframe. Users could also experiment more with \modelname (4.55 vs 2.6 prompt edits). Psychometric evaluation reinforced these findings, with 76.04\% positive responses and 83\% intending frequent use. Users praised \modelname's efficiency in data processing, component analysis, and model comparison. 
%While the error reinforcement component scored lower, with 25\% finding more errors than suggested, this feedback highlighted a desire for custom error checking capabilities (\S \ref{limitations}).

\begin{figure}[t]
    \centering
    \includegraphics[width=1\linewidth]{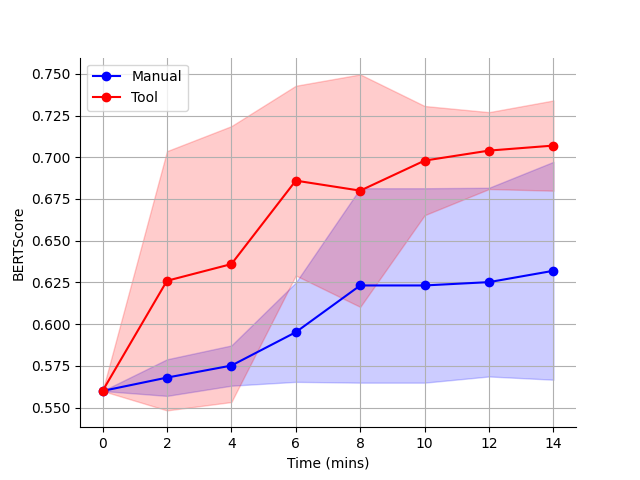}
    \caption{BERTScore Progression Over Time. Solid line is the average score while shaded region is the standard deviation.}
    %, demonstrating the tool's efficiency in improving output similarity during prompt migration.
    \label{fig:user_results}
\end{figure}

\begin{figure}[t]
    \centering
    \includegraphics[width=1\linewidth]{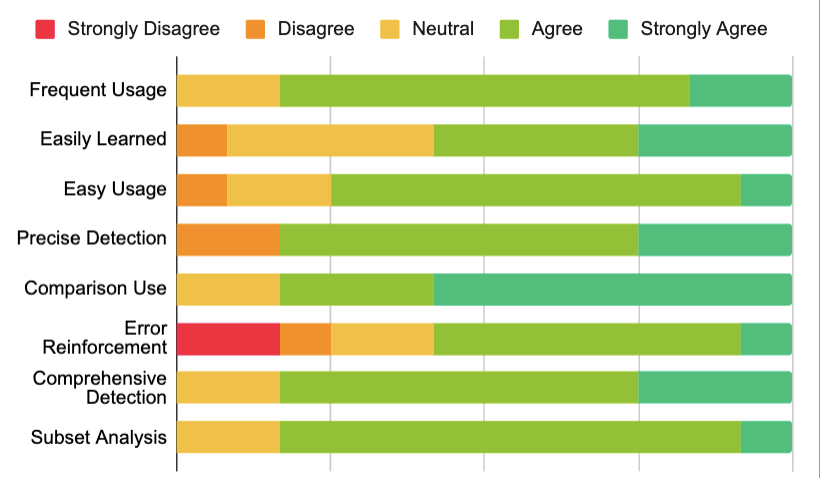}
    \caption{Post-Study Psychometric Evaluation Results. The x-axis labels are simplified for readability and the full questions are available in Section \S \ref{app:survey_questions}.}
    \label{fig:user_results}
\end{figure}

\section{Conclusion}
We present \modelname - a tool for regression testing guided LLM Migration. \modelname comprises of an interactive prompting interface tailored to regression testing needs, and an error discovery module that facilitates understanding differences in model outputs. The tools aims to help users in understanding where and why models score poorly on given metrics. Our user study indicated that the tool enables users identify twice as many errors, iterate with more prompt versions and achieve a higher score on evaluation metrics within the same time frame. We hope that our easy to setup, self-contained tool will facilitate broader adoption among those involved in LLM migration tasks.

%Through our user study \modelname helped users identify twice as many errors, experiment with 75\% more prompts versions and achieve a 12\% higher score on evaluation metrics within the same time frame. \\
% \input{sections/7_Limitations}

\bibliography{anthology,custom}
\bibliographystyle{acl_natbib}

\appendix
\newpage
\section{Additional Details}
\label{app:imp_details}

\paragraph{Tool Implementation Details}
\modelname is a web-based application.The entire tool was implemented using Python. For the user interface we used Reflex\footnote{https://reflex.dev/} while for the backend we made use of Langchain and litellm to query the various LLMs.
% Its preprocessing module loads a data file that stores the results from the AutoSxS libraries containing a list of prompts with response pairs and the ratings with rationales. Then it calls an LLM to summarize rationales into bullet points, generate cluster labels, and compute embeddings to be used for cluster assignments. The server, written in Python, loads this preprocessed data file and then transmits it to the client in JSON format. Once data is loaded into the client, all computations, such as filtering, sorting, and cluster assignments, are performed dynamically on web browser. The client-side code is written in TypeScript using the Lit webcomponents framework.3 When auser requests to regenerate rationale clusters, the server invokes calls to an LLM using a RPC call. "

\begin{table}[h!]
    \centering
    \small
    \begin{tabular}{p{1\linewidth}}
        \toprule
        \# \textbf{prompts...} \\
        prompts: \\
         ~~ - "Summarize this {{document}}"  \\
         ~~ - "Summarize this {{document}}, concisely and professionally:"  \\

        \# \textbf{models...} \\
        providers:  \\
          ~~ - openai:gpt-35-turbo-16k \\
          ~~ - meta-llama-3-8b \\
        \# \textbf{tests cases} \\
        tests: \\
        ~~ - vars: \\
             ~~~~ document: "file://docs.txt" \\
        ~~ assert: \\
             ~~~~~~ - type: bleu \\
             ~~~~~~ ~~   value: "Summary \dots" \\
             ~~~~~~ - type: bertscore \\
             ~~~~~~ ~~   value: "Summary \dots" \\
         \bottomrule
    \end{tabular}
    \caption{Example of a configuration file used to setup \modelname.}
    \label{tab:config_file}
\end{table}

\section{Error Discovery Implementation Details}
\label{app:error_discovery}
The goal-oriented approach has two prompts. The prompt used for generating the errors is in Table~\ref{tab:generator_prompt} and for selecting the model outputs in Table~\ref{tab:eval_prompt}. For the generator prompt, it is possible that for some instances all the model outputs might not fit into one prompt, hence we construct multiple prompts with different sets of samples so that GPT-4 can “see” all the different model outputs. We set temperature to be 0 for both the tasks. The baseline (non-goal oriented approach) used the prompt described in Table~\ref{tab:baseline_prompt}.\\
\par For generating the synthetic evaluation dataset, we use the following attributes set \textit{topic, writing style, stance, language, formatting, and country} and V was varied from 0.6 to 1.0. We prompted GPT-4 to generate the outputs.

\begin{table}[h!]
    \centering
    \small
    \begin{tabular}{p{1\linewidth}}
        \toprule
    Given two groups of inputs ( Group A and Group B ) and a Question, your task is to identify differences that make the groups different according to the specific question. Each input in a group starts with the token [ITEM]. \\
    Follow these guidelines: \\
    1. Only generate differences that help answer the question provided. \\
    2. Only generate 4-5 words description for each difference. \\
    3. Each difference description should start on a new line.  \\
    4. Each difference should be unique and relevant to the question provided. \\
    5. If there are no differences that make the groups different according to the question, output 'There are no differences that make the groups different according to the question provided'. \\
    Group A: \{\{Corpus A\}\} \\
    Group B: \{\{Corpus B\}\} \\
    Question: {{goal}} \\
    Compared to outputs in Group A, more outputs in Group B \\
         \bottomrule
    \end{tabular}
    \caption{Prompt used to generate the various errors as part of the goal oriented error discovery module.}
    \label{tab:generator_prompt}
\end{table}

\begin{table}[h!]
    \centering
    \small
    \begin{tabular}{p{1\linewidth}}
        \toprule
        Given two groups of outputs ( Model A and Models B ) and a Question, your task is to identify textual differences that answer the specific question. Each output in a model starts with the token [ITEM].  \\
        Follow these guidelines: \\
        1. Only generate differences that help answer the question provided. \\
        2. Only generate 4-5 words description for each difference.\\
        3. Each difference description should start on a new line. \\
        4. Each difference should be unique and should help answer the question provided. \\
        5. If there are no differences that make the groups different according to the question, output 'There are no differences that make the groups different according to the question provided'. \\
        Model A Outputs: \{\{Corpus A\}\}\\
        Model B Outputs: \{\{Corpus B\}\} \\
        Question: \{\{goal\}\} \\
        To answer the question, we can see that, compared to outputs from Model A, more outputs from Model B are \\
         \bottomrule
    \end{tabular}
    \caption{Prompt used to select the various model outputs which contain a given error type.}
    \label{tab:eval_prompt}
\end{table}

\begin{table}[h]
    \centering
    \small
    \begin{tabular}{p{1\linewidth}}
        \toprule
        Given two groups of inputs ( Group A and Group B ), identify all stylistic, syntactic and semantic differences that make the groups different. Some possible differences could be common words, phrases, or patterns in writing style that are present in one group but not in the other group. Each input in a group starts with the token [ITEM]. Only generate 4-5 words description for each difference, and each difference description should start on a new line. Ensure to cover all the above 3 categories of differences. Do not output descriptions that start with words like 'In Group A' or 'Group B ..'.\\
Group A: \{\{set_a\}\}\\
Group B: \{\{set_b\}\}\\
Compared to outputs in Group A, majority outputs in Group B \\
         \bottomrule
    \end{tabular}
    \caption{Prompt used as a baseline to find differences between two groups. This is a standalone, non-goal-oriented prompt}
    \label{tab:baseline_prompt}
\end{table}

\begin{figure*}[t]
    \centering
    \includegraphics[width=1\linewidth]{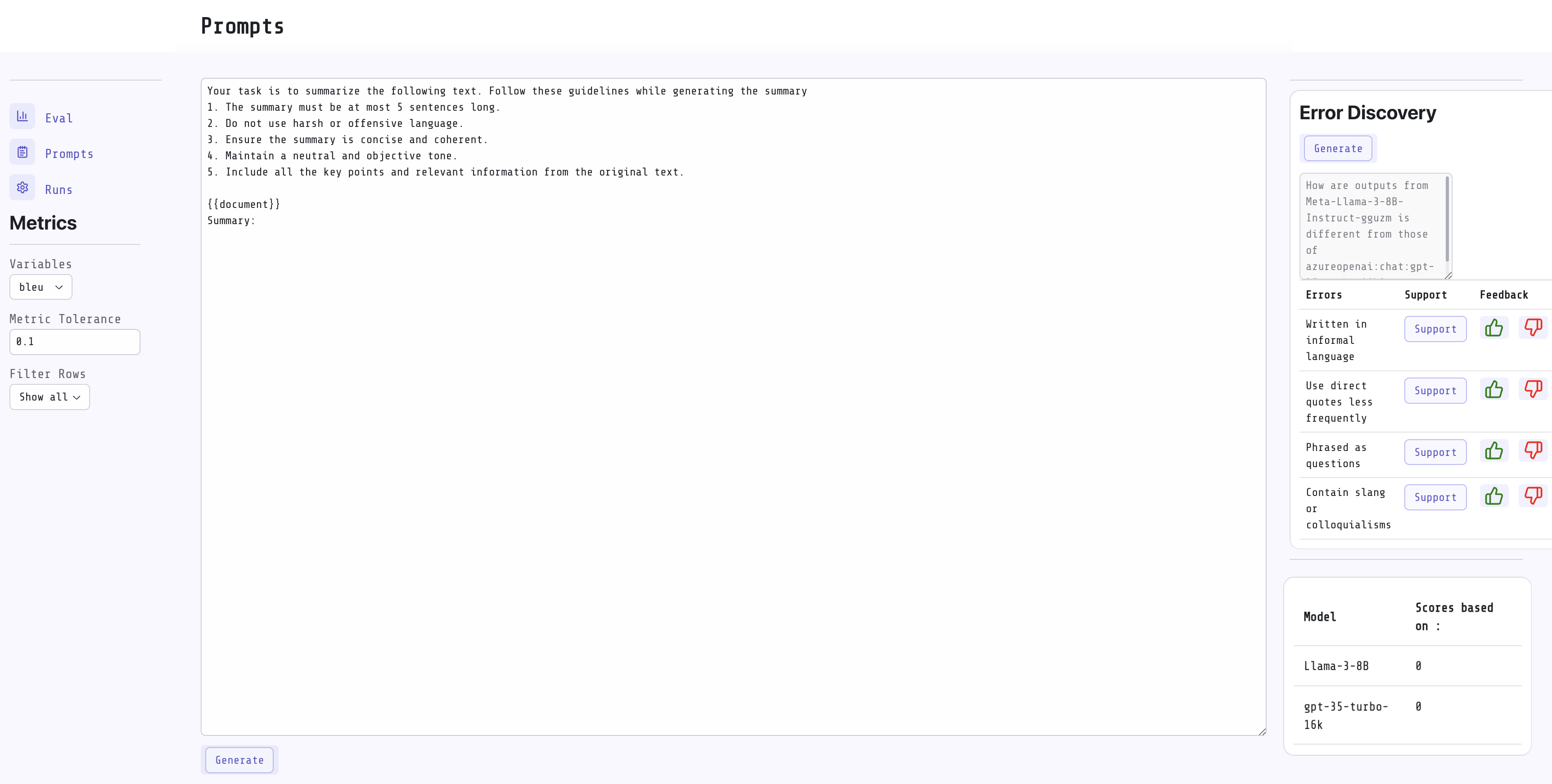}
    \caption{Prompts Page: The user can edit/update the model prompts using the Prompts tab. }
    \label{fig:prompts_page}
\end{figure*}

\begin{figure*}[t]
    \centering
    \includegraphics[width=1\linewidth]{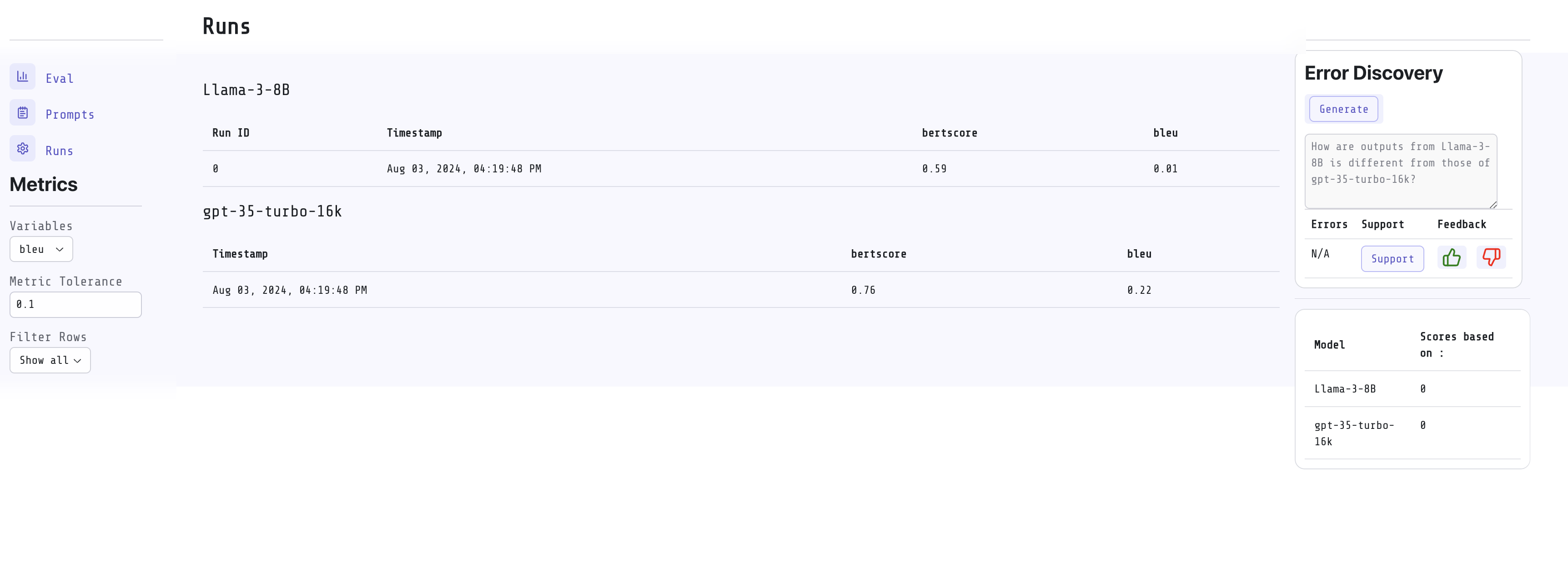}
    \caption{Runs Page: This page provides a tabular visualization of the various prompt versions.}
    \label{fig:runs_page}
\end{figure*}

\section{User Study Details}
\label{app:user_study}

\subsection{Participant Recruitment}
We recruited 12 participants for this study, each with at least two years of experience in ML Engineering or prompt engineering with LLMs. All participants were ML Engineers or Research Scientists from industrial settings, regularly working with LLMs for task-oriented use cases. Recruitment was conducted via an internal messaging service, dissemintated to individuals who had no conflicting interest. 

Participants were selected based on their expertise to ensure informed feedback on the LLM Migration tool. All interviews were conducted in person. Compensation included a single-meal voucher or gift of equivalent value in California. 

\subsection{Post User Survey Questions}
\label{app:survey_questions}

\begin{itemize}
    \item I think I would like to use this system frequently.
    \item I would imagine that most people would learn to use this system very quickly.
    \item I found the system very easy to use.
    \item The error discovery module helped me identify errors quickly.
    \item This tool could be useful for comparing two LLMs.
    \item The error discovery module helped reinforce the errors I had observed.
    \item The error discovery module helped me to quickly identify the data points with the a common error. 
    \item The tool provided support to analyze different subsets of the data according to the user needs. 
\end{itemize}

\end{document}